\documentclass[10pt]{article}

\usepackage{amssymb}
\usepackage{amsfonts}
\usepackage{amsmath}
\usepackage{amscd}
\usepackage[all,cmtip]{xy}
\usepackage{amsthm}
\usepackage{setspace}
\usepackage{enumerate}
\usepackage{float}
\usepackage{tikz}

\usetikzlibrary{decorations.pathmorphing}

\usetikzlibrary{patterns}

\theoremstyle{plain}

\newtheorem*{thm*}{Theorem}
\newtheorem*{NLL}{Arithmetic NLL}
\newtheorem*{GNLL}{Geometric NLL}

\theoremstyle{definition}

\newcommand{\CC}{{\mathbb C}}

\newcommand{\QQ}{{\mathbb Q}}
\newcommand{\FF}{{\mathbb F}}
\newcommand{\ZZ}{{\mathbb Z}}
\newcommand{\PP}{{\mathbb P}}
\renewcommand{\AA}{{\mathbb A}}

\newcommand{\ol}{\overline}

\newcommand{\Gal}{{\rm Gal}}

\newcommand{\GL}{{\rm GL}}

\title{Local Langlands Duality and a Duality of Conformal Field Theories}
\author{Martin T. Luu}

\date{}

\begin{document}

\newcommand{\Addresses}{{
\bigskip
\footnotesize

M. Luu, \textsc{Department of Mathematics, University of Illinois at Urbana-Champaign, IL 61801, USA} \par \nopagebreak \textit{E-mail address:} \texttt{mluu@illinois.edu}

}}

\maketitle

\thispagestyle{empty}

\noindent

\begin{abstract}
We show that the numerical local Langlands duality for $\GL_{n}$ and the T -- duality of two-dimensional quantum gravity arise from one and the same symmetry principle. The unifying theme is that the local Fourier transform in both its $\ell$-adic and complex incarnation gives rise to symmetries of arithmetic and geometric local Langlands parameters. 
\end{abstract}

\section{Introduction}
In the 1990's physicists gave rigorous mathematical proofs of the T -- duality, also known as p -- q duality, of 2D quantum gravity, see \cite{FKN} and \cite{KM}. Recently, we revisited this topic in \cite{LUU} and in joint work with Albert S. Schwarz \cite{LUU2}, giving a proof of this duality based on the complex local Fourier transform of Bloch-Esnault \cite{BE} and Lopez \cite{LOP}. This transform has an $\ell$-adic analogue and it is then natural to ask if there is arithmetic meaning to this new proof of the T -- duality. In the present work we explain that indeed there is such meaning and that from a certain perspective the arithmetic analogue is the numerical local Langlands duality for $\GL_{n}$ over local fields.

To describe this passage between physics and arithmetic one should move along the four pillars of Weil's augmented Rosetta stone, see \cite{FRE}: These pillars are number fields, curves over finite fields, Riemann surfaces, and quantum physics. The analogies are particularly interesting when comparing various dualities that exist within the four different frameworks. Before explaining our results concerning this relation of dualities we recall a global analogue of the passage. 

Consider first the arithmetic duality given by the relation between Galois representations and automorphic representations. One can say that the study of this duality started, in its true non-abelian generality, in the work of Eichler \cite{EIC}. By now, it is a well studied theme and often called a Langlands duality. For example in the context of $\GL_{n,\QQ}$ it is a popular conjecture, see for example \cite{TAY}, that every irreducible representation
$$\rho : \Gal(\ol{\QQ}/\QQ) \longrightarrow \GL_{n}(\ol{\QQ}_{\ell})$$
that is continuous for the $\ell$-adic topology on the target and the profinite topology on the source and that is unramified at all but finitely many places and de Rham at $\ell$ is attached to a cuspidal automorphic representation of $\GL_{n,\QQ}$. This has a global geometric analogue:

Given a Riemann surface $X$, the unramified geometric Langlands duality for $\GL_{n}$ associates to every flat $\GL_{n}$-bundle over $X$ a certain D-module on the moduli stack of $\GL_{n}$-bundles over $X$. The fourth aspect of the Rosetta stone, quantum physics, enters through the work of Kapustin and Witten \cite{KW} that relates the geometric Langlands duality to the S -- duality of 4D gauge theories.

We will show that in analogy to the above global passage between arithmetic and physics, the T -- duality of 2D quantum gravity can be related to dualities in the local incarnations of the first three pillars of the Rosetta stone. In the first two pillars, the numerical local Langlands correspondence for $\GL_{n}$ over local fields has been obtained in work of Henniart \cite{HEN}. To move to the third pillar we introduce a local geometric numerical Langlands correspondence. Afterwards, to move to the fourth pillar, we explain how quantum physics enters the picture: 

We describe how one can associate Virasoro constrained $\tau$-functions of the KP hierarchy to suitable local geometric Langlands parameters. These $\tau$-functions are relevant in quantum field theory, in particular in 2D quantum gravity where the partition functions of the various models of the theory can be described in terms of $\tau$-functions. The question then occurs whether there is an analogue of the local numerical Langlands correspondence in these physical theories. It turns out that, rightly interpreted, there indeed is and it has the same underlying mechanism as the arithmetic duality, namely symmetries of local Langlands parameters coming from the local Fourier transform. We explain how this viewpoint yields the relation between the Langlands duality and the T -- duality.

\section{Numerical local Langlands: Arithmetic and Geometry}
Up to this point, the numerical local Langlands correspondence for $\GL_{n}$ has been known for the local incarnations of the first two pillars of Weil's Rosetta stone, namely fields of the form $\FF_{q}(\!(t)\!)$ and finite extension of $\QQ_{p}$. 

Fix a non-archimedean local field $K$ with finite residue field of size $q$. Let 
$\mathcal A^{0}(n)$ denote the set of isomorphism classes of supercuspidal representations of $\textrm{GL}_{n}(K)$
and let $\mathcal G^{0}(n)$ denote the set of isomorphism classes of $n$-dimensional Weil-Deligne representations $(r,N)$ of $W_{K}$ with $r$ irreducible. The local Langlands correspondence predicts a bijection between $\mathcal A^{0}(n)$ and $\mathcal G^{0}(n)$ that preserves interesting arithmetic data such as the conductor. A natural approach to test this bijection is to put more and more constraints on both sets, until one obtains finite sets. These should then have the same size and this is the idea of the numerical local Langlands correspondence.

This correspondence plays a crucial role in the original proofs of the local Langlands correspondence for $\GL_{n}$ over a $p$-adic field by Harris-Taylor \cite{HT} and Henniart \cite{HEN}: It reduces the task of constructing a suitable bijection between the relevant Weil-Deligne representations and smooth representations of $\GL_{n}$ to the task of constructing a suitable injection. This is a significant simplification. Consider for example the proof by Harris and Taylor:

Let $K$ be a $p$-adic field and suppose given a smooth admissible representation $\tilde \pi$ of $\GL_{n}(K)$. Representations whose matrix coefficients are analytically nice, in particular supercuspidal representations, can be realized thanks to work of Clozel as local components of suitable automorphic representations. This means that there exists a suitable global object, an automorphic representation $\pi$ over a global field $F$, which factors as a restricted tensor product 
$$\pi \cong \otimes'_{v} \pi_{v}$$
such that there is a finite place $v_{0}$ of $F$ such that $F_{v_{0}} \cong K$ and $\pi_{v_{0}} \cong \tilde \pi$. The candidate for the local Langlands correspondence is then given by
$$\tilde  \pi \mapsto \rho_{\pi,\ell}|_{\textrm{Gal}(\ol{F}_{v_{0}} / F_{v_{0}})}$$
where $\rho_{\pi,\ell}$ is a suitable $\ell$-adic representation attached to $\pi$. However, it is not clear from the construction what the image of this correspondence is. The great use of the numerical local Langlands correspondence is that it makes it unnecessary to understand the  image: Henniart \cite{HEN} used the numerical correspondence to show that any injection 
$$\mathcal G^{0}(n) \hookrightarrow \mathcal A^{0}(n)$$
that is compatible with twists by unramified characters and preserves conductors has to be a bijection. We now recall the key features of the arithmetic theory before describing a geometric analogue.

\subsection{The arithmetic theory}
Let $K$ be again a non-archimedean local field with residue field $\FF_{q}$. Define the set
$$C(n,j)_{\textrm{arith}}=\bigcup_{d | n} \; \bigcup_{ k \cdot \frac{n}{d} \le j} \frac{ \Big \{ \sigma=(r,N) \in \mathcal G^{0}(d) \; \Big | \;  r|_{I_{K}} \textrm{ is irreducible and } \textrm{sw}(\sigma)=k \Big \}}{ \textrm{twist by unramified characters}}$$
where $\textrm{sw}(-)$ denotes the Swan conductor and $I_{K}$ is the inertia subgroup of $W_{K}$. This is known to be a finite set by work of Koch and hence it makes sense to define the number
$$c(n,j)_{\textrm{arith}}:= | C(n,j)_{\textrm{arith}}|$$
The following was conjectured by Koch, in a slightly different but equivalent form:
\begin{NLL}
One has
$$c(n,j)_{\textrm{\emph{arith}}}=(q-1)q^{j}$$
with $q$ the size of residue field of $K$.
\end{NLL}
This is one of several ways to formulate the numerical local Langlands correspondence. See for example \cite{LOR} for background and history concerning this correspondence. After long and laborious calculations, Koch \cite{KOC} was able to prove the conjecture if the residue characteristic $p$ of $\FF_{q}$ divides $n$ at most linearly. Henniart used completely different methods in \cite{HEN} to prove the general case. This shift in the approach to the conjecture involves the local Fourier functors and is crucial for us to make the connection with quantum duality.

Our aim is now to obtain a numerical local geometric Langlands correspondence. We start by describing arithmetic and geometric local Langlands parameters in a unified manner.

\begin{itemize}
\item
Let $K$ be a non-archimedean local field and let $W_{K}$ be the corresponding Weil group. An arithmetic local Langlands parameter $\mathcal L$ for $\GL_{n}$ with trivial monodromy operator is a pair $(V,\rho)$ where $V$ is an $n$-dimensional $\CC$-vector space and $\rho$ is a homomorphism
$$\rho : W_{K} \longrightarrow \GL(V)$$
with open kernel. 
\item
A geometric local Langlands parameter $\mathcal L$ for $\GL_{n}$ is simply a connection $(V,\nabla)$ on the formal punctured disc, see \cite{FG}. After choosing a local coordinate, this consists of the data of an $n$-dimensional $\CC(\!(t)\!)$-vector space $M$ together with a $\CC$-linear map 
$$\nabla : M \longrightarrow M$$
such that
$$\nabla(f \cdot m) = f \cdot \nabla(m) + \frac{\textrm{d} f}{\textrm{d} t} \cdot m$$
for all $f \in \CC(\!(t)\!)$ and all $m \in M$. 
\end{itemize}
There will be two important integer parameters $n$ and $j$ associated to such arithmetic and geometric Langlands parameters:

For an arithmetic Langlands parameter $\mathcal L= (V,\rho)$ let
$$n(\mathcal L) := \dim_{\CC}V \; \; \textrm{ and } \; \;
j(\mathcal L)=\textrm{sw}(\rho)$$
where $\textrm{sw}(-)$ is the Swan conductor. For a geometric Langlands parameter $\mathcal L=(\nabla, V)$ define
$$n(\mathcal L) := \dim_{\CC(\!(t)\!)}V \;\; \textrm{ and } \;\;
j(\mathcal L)= \sum \textrm{slopes of }\nabla$$
The integer $j(\mathcal L)$ is called the irregularity of the connection. We will give a concrete description later on in the proof of the geometric numerical local Langlands correspondence. 

For $n, j \in \ZZ^{> 0}$ let $\textrm{LL}_{\textrm{arith}}(n,j)$ denote the set of isomorphism classes of arithmetic local Langlands parameters of dimension $n$ and Swan conductor $j$, and let $\textrm{LL}_{\textrm{arith}}^{\circ}(n,j)$ denote the subset of irreducible parameters. The geometric analogue is the set $\textrm{LL}_{\textrm{geom}}(n,j)$ of isomorphism classes of irreducible geometric local Langlands dimension $n$ and irregularity $j$ and the subset $\textrm{LL}_{\textrm{geom}}^{\circ}(n,j)$ consisting of irreducible connections.

\subsection{A geometric version}
Our aim is now to find a set $C(n,j)_{\textrm{geom}}$ of isomorphism classes of local geometric Langlands parameters that is defined in a similar manner to $C(n,j)_{\textrm{arith}}$ and whose suitably interpreted size corresponds to $c(n,j)_{\textrm{arith}}$. A priori, there might not be much reason to expect the existence of such a geometric analogue, but we will show the existence of this analogue if one does not blindly translate all the arithmetic notions to the geometric setting. Rather, one should take a cue from the representation theory of $\GL_{n}$ over $p$-adic fields: 

 The origin for the summation over $d|n$ and over suitable $k$'s in the definition of $C(n,j)_{\textrm{arith}}$ is that this is a count of suitable discrete series representations rather than supercuspidal representations. The reason that one can count these more easily is that these representations are all on the same footing when viewed via the local Jacquet-Langlands correspondence as representations of a suitable inner form of the general linear group. In the arithmetic setting it is thus a fact that the ``discrete series count'' gives a simpler formula than the ``supercuspidal count'' where
\begin{eqnarray*}
\textrm{ discrete series count: } \textrm{ all pairs $(d,k)$ such that }  &d \big | n&  \;\;\; \textrm{ and } \;\;\; k \le j  \cdot \frac{d}{n}\\\\
\textrm{ supercuspidal count: }  \textrm{ all pairs $(d,k)$ such that } &d=n&  \;\;\; \textrm{ and } \;\;\; k \le j
\end{eqnarray*}
Under the analogies of Weil's Rosetta stone, the geometric residue field $k_{g} := \CC$ of $\CC(\!(t)\!)$ corresponds to the arithmetic residue field $k_{a}:=\FF_{q}$ of $K$. Hence one might expect that a geometric analogue of $\CC(n,j)_{\textrm{arith}}$ should be described via suitable parameters in the residue field $\CC$. Furthermore,
given the formula
$$c(n,j)_{\textrm{arith}}=(q-1)q^{j}$$
it is natural to try to relate this to a point count of a variety over a finite field. It seems unknown if there is indeed a variety whose points count the set $C(n,j)_{\textrm{arith}}$ in a natural manner. However, after the fact, meaning with the above formula in hand, the arithmetic numerical local Langlands correspondence can be phrased as the existence of a bijection
$$\xymatrix{ C(n,j)_{\textrm{arith}} \;\; \ar[rr]^-{1:1} & & \;\; \left(\AA^{j} \times \mathbb{G}_{m} \right )(k_{a}) \ar[ll] }$$
The occurrence of this variety might seem artificial but something interesting occurs when moving to the geometric pillar of the Rosetta stone: 

Up to avoiding a hypersurface related to reducibility of local geometric Langlands parameters, the same variety occurs in a natural analogue of the numerical local Langlands duality in the geometric set-up:
\begin{GNLL}
There is a set $C(n,j)_{\textrm{\emph{geom}}}$ of local geometric Langlands parameters for $\GL_{n}$, whose definition is modeled on the supercuspidal count, such that there is a bijection
$$\xymatrix{C(n,j)_{\textrm{\emph{geom}}}  \;\; \ar[rr]^-{1:1} && \;\;  \left ( U \times \mathbb{G}_{m} \right )(k_{g}) \ar[ll]}$$
for the open subspace $U$ of $\AA^{j}=\textrm{\emph{Spec }} \CC[x_{1},\cdots,x_{j}]$ given by complement of the hypersurface defined by the equation $\prod_{i: (i,n)=1} x_{i} = 0$. 
\end{GNLL}
\begin{proof}
We start by recalling how the classical results of Levelt and Turrittin allow to describe irreducible $n$-dimensional connections on the formal punctured disc:

Given a Laurent series $f=\sum a_{i} t^{i} \in \CC(\!(t^{1/n})\!)$, consider the $n$-dimensional connection $E_{f}$ over $\CC(\!(t)\!)$ given by the vector space $\CC(\!(t^{1/n})\!)$ with connection
$$\frac{\textrm{d}}{\textrm{d}t} +\frac{f(t)}{t}$$ 
One can describe the connection using slightly different language: Given a map 
$\rho : \CC[\![t]\!] \longrightarrow \CC[\![u]\!]$
which takes $t$ to some element in $u \CC[\![u]\!]$ one can define associated push-forward and pull-back operations on the categories of connections on the formal punctured disc with local coordinate $t$ and $u$ respectively. For $i \in \ZZ^{\ge 1}$, denote by $[i]$ the map $\rho$ that takes $t$ to $u^{i}$. Then 
$$E_{f} \cong  [n]_{*} \left (\CC(\!(t^{1/n})\!), \frac{\textrm{d}}{\textrm{d}t^{1/n}}+nt^{(n-1)/n}\frac{f}{t^{1/n}} \right )$$
This connection is irreducible if and only if $f$ is not in $\CC(\!(t^{1/m})\!)$ for some $0 < m <n$. The Levelt-Turrittin classification then implies that every irreducible $n$-dimensional connection is isomorphic to some such $E_{f}$, and $f$ is unique up to adding an arbitrary element of
$$\frac{1}{n}\ZZ + t^{1/n}\CC[\![t^{1/n}]\!]$$
The irregularity of such an irreducible connection $E_{f}$ for
$$f= a_{-j} t^{-j/n} + \textrm{higher order terms}$$
is $j$. One observes immediately a crucial difference between the arithmetic and geometric side: 

In the latter, all irreducible parameters are obtained from one-dimensional ones via push-forward. In the former, the question of which parameters of $W_{K}$ are obtained by induction from a parameter of $W_{K'}$ for some finite extension $K'/K$ is subtle. Indeed, for example in the local Langlands correspondence for $\GL_{2}$ over $p$-adic fields, a crucial case is $\GL_{2}(\QQ_{2})$, precisely since in this case there exist non-induced parameters. 

Consider for example for an odd $c \ge 3$ the set of isomorphism classes of Weil-Deligne representations of $W_{K}$ that are induced from a proper subgroup of $W_{K}$ and that have Artin conductor $c$ and trivial determinant on a choice of uniformizer. Here $K$ is a $p$-adic field and we denote by $q$ the size of the residue field of $K$. Work of Tunnell \cite{TUN} shows that this set has size
$$2(q-1)^{2}q^{c-3}(1-X(c))q^{-[(c+1)/6]}$$
where the function $X(c)$ depends on how $c$ compares to $6 \cdot \textrm{val}_{K}(2)+1$ and also on the congruence of $c$ modulo $3$. As an application of the local Jacquet-Langlands correspondence one sees that there is no such subtlety for the smooth representations of $\GL_{2}$ over $p$-adic fields. It follows that on the Galois side the number of local Langlands parameters that are not induced also has a subtle behavior that offsets the subtle count of induced parameters. 

Since there is no dichotomy between induced and non-induced parameters on the geometric side, we take this as our first indication that some liberty should be taken when transferring the arithmetic statements to local geometric Langlands parameters.

As previously indicated, it turns out that the right analogue of the discrete series count in the geometric setting is the supercuspidal count. Hence we set $d=n$ when defining the geometric analogue $C(n,j)_{\textrm{geom}}$ of $C(n,j)_{\textrm{arith}}$. We now discuss geometric analogues of the other constraints involved in the definition of $C(n,j)_{\textrm{arith}}$. 

First consider the process of identifying representations that differ by a twist by an unramified character. A natural geometric analogue is the process of tensoring a connection with a holomorphic one-dimensional connection. However, this does not change the isomorphism class of the connection. Hence, given an irreducible connection $E_{f}$ we can reduce to the case where $f$ is of the form
$$f=\sum_{i\le 0} a_{i} z^{i/d}$$
The condition that the Swan conductor is $k$ corresponds to the irregularity being $k$, hence the relevant $f$'s are of the form
$$f=\sum_{-k \le  i\le 0} a_{i} z^{i/d}$$
Given a Weil-Deligne representation $(r,N)$, the condition that $r$ restricted to inertia is irreducible we simply translate to the condition that $E_{f}$ is irreducible. This means that there is $i$ with $a_{i}\ne0$ and $\textrm{gcd}(i,d)=1$. With all these conventions we obtain the geometric analogue of $C(n,j)_{\textrm{arith}}$ as
$$C(n,j)_{\textrm{geom}} := \bigcup_{k\le j} \textrm{LL}_{\textrm{geom}}^{\circ}(n,k)$$
Due to the Levelt-Turrittin classification and the isomorphism
$\xymatrix{\exp(2 \pi \sqrt{-1} n - ) : \CC/ \frac{1}{n}\ZZ  \ar[r]^-{\sim} & \CC^{*}}$, there is a bijection
$$C(n,j)_{\textrm{geom}} \longleftrightarrow \Big \{(a_{1},\cdots, a_{j}) \in \CC^{j} \; \big | \;   \prod_{i: (i,n)=1} a_{i} \ne 0 \Big \} \times \mathbb{G}_{m}(k_{g})$$
and hence the claimed version of a geometric numerical local Langlands correspondence follows. The similarity between this set and its arithmetic analogue in particular gives a complex shadow of the structure of the abelian part of the inertia group of a local field.
\end{proof}

\section{Quantum Physics}
Due to the work of Henniart and Laumon \cite{HEN}, \cite{LAU} in conjunction with the discussion of the previous section, it can be said that there is a numerical local Langlands duality for the first three pillars of the Rosetta stone. The question then arises whether there exists a version of this duality in the fourth pillar, quantum physics. Furthermore, if it does, what is its relation to known dualities in quantum field theory? 

As a first step towards our answer, we now explain how quantum physics enters the picture at all, forgetting for now about the desired duality. As a particular case we discuss  how one can describe 2D quantum gravity via local geometric Langlands parameters. This is not an obvious consequence of the original formulations of this quantum field theory but its importance has recently been stressed in \cite{LUU}, \cite{LUU2}, \cite{SCH}.

The crucial point is that as a consequence of the Levelt-Turrittin theory one can attach to irreducible connections on the punctured disc special $\tau$-functions of the KP-hierarchy. We now explain this. The KP-hierarchy of partial differential equations can be described in the following manner:

Consider a Lax operator
$$L(t_{1},t_{2},\cdots)=\partial_{t_{1}} + u_{-1}(t_{1},t_{2},\cdots) \partial_{t_{1}}^{-1}+\cdots$$
depending on the infinite set $t_{1},\cdots$ of time variables. This is a pseudo-differential operator and the Lax equations are given by
$$\frac{\partial }{\partial t_{i}}L=[L^{i}_{+},L]$$
where $L^{i}_{+}$ denotes the part of $L^{i}$ that does not involve negative powers of $\partial_{t_{1}}$. This yields constraints for the coefficients $u_{-1}, u_{-2}, \cdots$. For example, $u_{-1}$ has to satisfy the KP equation
$$\partial_{t_{1}}\left( 4 \partial_{t_{3}} u_{-1} -12 u_{-1} \partial_{t_{1}} u_{-1} - \partial^{3}_{t_{1}} u_{-1} \right) = 3 \partial^{2}_{t_{2}} u_{-1}$$
This type of equation was introduced in the theory of water waves and is a generalization in one higher dimension of the KdV equation. A $\tau$-function of the hierarchy allows to describe the coefficients of the Lax operator in a concise manner, one has for example
$$u_{-1}=\partial_{t_{1}}^{2} \ln \tau $$
$$u_{-2}=\frac{1}{2}(\partial_{t_{1}}^{3}+\partial_{t_{1}}\partial_{t_{3}}) \ln \tau $$
$$\vdots$$
The relation with quantum physics comes from the fact that partition functions of certain quantum field theories are expected to be expressible in terms of special KP $\tau$-functions. For example, the work of Kontsevich shows that the partition function of topological gravity is the square of a $2$-reduced KP $\tau$-function that satisfies certain Virasoro constraints:

The Virasoro algebra is given by
$\textrm{Vir}:=  \langle L_{n} \rangle_{n\in \ZZ} \oplus \langle c\rangle$
with $c$ a central element and
$$ [L_{m},L_{n}]=(m-n)L_{n+m} + \frac{m^{3}-m}{12}\delta_{m+n,0} \cdot c$$
Consider the differential operators
$$\mathcal L_{n}= \frac{1}{2}\sum_{i+j=-n}  ij t_{i} t_{j} + \sum_{i-j=-n} it_{i} \partial_{j} + \frac{1}{2} \sum_{i+j=n} \partial_{i}\partial_{j} $$
These yield a representation of the Virasoro algebra with central charge $1$ on the space $\CC[\![t_{1},t_{2},\cdots]\!]$. In particular, this yields an action on the $\tau$-functions of the KP hierarchy.

By a Virasoro constrained $\tau$-function $\tau(t_{1},t_{2},\cdots)$ of the KP hierarchy we mean a $\tau$-function annihilated by suitable operators $\mathcal L_{n}$ as above. The following discussion describes how one can attach special points of the Sato Grassmannian to suitable local geometric Langlands parameters and hence via Sato's work, see for example \cite{MUL}, one can attach Virasoro constrained $\tau$-functions to these parameters. As part of this construction one obtains KP dynamics on the set of Langlands parameters by letting the corresponding points of the Grassmannian flow along the KP flows. This is important in the duality results that we describe in the next section.

Suppose given an irreducible connection
$$\mathcal L= (\CC(\!(t)\!)^{n}, \frac{\textrm{d}}{\textrm{dt}} + A) \in \textrm{LL}_{\textrm{geom}}^{\circ}(n,j)$$
for $A \in \mathfrak g \mathfrak l_{n}(\CC(\!(t)\!))$. By the Levelt-Turrittin classification it follows that after extending scalars from $\CC(\!(t)\!)$ to $\CC(\!(t^{1/n})\!)$ one can gauge transform $A$ into diagonal connection matrix $D$. Furthermore, it is known that this can be obtained by a gauge transformation of the form
$g \in  M_{n}(\CC)[\![1/z]\!]$. 
The beautiful observation described in much more detail by Schwarz in \cite{SCH} is that each column $\textbf{u}$ of $g$ gives rise to a Virasoro constrained KP $\tau$-function in the following manner. The equation
$$g^{-1}Ag + g^{-1}\frac{\textrm{d}}{\textrm{d}t}g  = \begin{bmatrix}
d_{1} & & &\\
&\ddots &&\\
&&d_{n}
\end{bmatrix}$$
implies that
$$\frac{\textrm{d}}{\textrm{d}t} \textbf{u} + \lambda(z) \textbf{u} = A \textbf{u}$$
for some Laurent series $\lambda(z)$. Let $u_{i}$ denote the $i$'th component of $\textbf{u}$ and for $0 \le i \le n-1$ let 
$$v_{i}=z^{i}u_{i}= z^{i} + \textrm{ lower order terms } \in \CC(\!(1/z)\!)$$
Now define a $\CC$-subspace of  $\CC(\!(1/z)\!)$ by
$$V:= \textrm{span}_{\CC} \Big \{ z^{pj}v_{i} \; \Big | 0\le i \le p-1, j \ge 0 \Big \}$$
Let $Gr$ be the big-cell of the Sato Grassmannian. It is the set of 
complex subspaces of $\CC(\!(1/z)\!)$ whose projection onto $
\CC[z]$ is an isomorphism. In particular, the above constructed 
space $V$ is a point of this big cell. By results of Sato, to every such 
point one has an associated $\tau$-function of the KP hierarchy. 

The point $V$ satisfies
$$
z^{n} V  \subseteq  V $$
$$
\left( \frac{1}{nz^{n-1}}\cdot \frac{\textrm{d}}{\textrm{d} z} +\frac{1-n}{2n} \frac{1}{z^{n}}  - \sum c_{i} z^{i} \right )V \subseteq  V$$
where the $c_{i}$'s can be explicitly described in terms of the $d_{i}$'s. By applying the Boson-Fermion correspondence this translates to Virasoro constraints for the associated KP $\tau$-function, see for example \cite{FKN2}.

One obtains a map
$$\Phi : \bigcup_{n,j} \textrm{LL}_{\textrm{geom}}^{\circ}(n,j) \longrightarrow \Big \{\textrm{Virasoro constrained $\tau$-functions of the KP hierarchy} \Big \}$$
such that for each geometric Langlands parameter $\mathcal L \in \textrm{LL}_{\textrm{geom}}^{\circ}(n,j)$ there exist constants $c_{i}$ such that
$$\partial_{t_{n}}\Phi(\mathcal L) = 0$$
$$\left(\mathcal L_{n} + \sum_{i \le j} c_{i}\partial_{t_{i}} \right )  \Phi(\mathcal L) = 0$$
Finally one arrives at quantum physics: 

Let $p$ and $q$ be two positive co-prime integers. For every ordered pair $(p,q)$ there is a corresponding model of 2D quantum gravity, see for example \cite{FKN}, with associated partition function denoted by 
$\textrm{Z}_{p,q}(t_{1}, \cdots, t_{p+q})$. On the level of rigor of physics, it is defined to be
$$\textrm{Z}_{p,q}(t_{1}, \cdots, t_{p+q})= \sum_{h} \int_{g} \exp(\dots)$$
where the exponential term is derived from the equations of general relativity and one sums over all possible values of the genus $h$ of the surface and integrates over all metrics $g$. This is crucial since in quantum gravity the metric is supposed to become dynamical. 

It is a crucial insight that the partition function satisfies 
$$Z_{p,q}(t_{1},\cdots)=\tau_{p,q}^{2}(t_{1},\cdots)$$
where $\tau_{p,q}$ is a $\tau$-function of the KP hierarchy that satisfies certain Virasoro constraints. On the level of the Sato Grassmannian $\tau_{p,q}$ can be described by a point $W_{p,q}(t_{1},\cdots,t_{p+q})$ of $Gr$ such that
$$
z^{p} W_{p,q}(t_{1},\cdots,t_{p+q}) \subseteq W_{p,q}(t_{1},\cdots,t_{p+q})
$$
$$
\left(\frac{1}{pz^{p-1}}\frac{\textrm{d}}{\textrm{d}z} + \frac{1-p}{2p}\frac{1}{z^{p}}- \frac{1}{p} \sum_{i=1}^{p+q} i t_{i} z^{i-p} \right) W_{p,q}(t_{1},\cdots,t_{p+q}) \subseteq W_{p,q}(t_{1},\cdots,t_{p+q})$$
In fact, one can write down the local geometric Langlands parameter that is associated to the $(p,q)$ model of 2D quantum gravity via the above described passage. It is given by
$$\mathcal L_{p,q}:=[p]_{*}\left( (\CC(\!(z)\!), \frac{\textrm{d}}{\textrm{d}z} + pz^{p-1}(\frac{1-p}{2p}\frac{1}{z^{p}}- \frac{1}{p} \sum_{i=1}^{p+q} i t_{i} z^{i-p})) \right )$$
It is shown in \cite{SCH2} that these points can be constructed from KP deformations of the quantization of the pair $(\partial_{x}^{p},\partial_{x}^{q})$ of commuting differential operators.

\section{Duality: Symmetries of Langlands parameters}
In the previous section it was explained how the fourth pillar of Weil's Rosetta stone enters the picture through Virasoro constrained $\tau$-functions of the KP hierarchy. We now explain that from this point of view, the numerical local Langlands correspondence does indeed correspond to a quantum duality, namely to the well studied T -- duality of 2D quantum gravity. 

We refer to \cite{FKN} for a detailed description of this quantum field theory and we now focus on a description of the duality of the theory. The starting point is the matter content of the theory before gravity is introduced into the picture. The matter content is chosen to be a minimal model conformal field theory. It has central charge
$$c(p,q)=1- 6 \frac{(p-q)^{2}}{pq}$$
for positive co-prime integers $p$ and $q$. Note that under this constraint on $p$ and $q$ the value of $c(p,q)$ determines $p$ and $q$ up to exchanging $p$ and $q$. It is known that there is an associated rational conformal field theory. To introduce gravity one couples this $(p,q)$ minimal model, where the order of $p$ and $q$ does not matter, to Liouville gravity. 

To explain the duality there is now a crucial observation. Let $\Sigma$ be a surface with a metric $g^{ij}$ of scalar curvature $R$. The un-renormalized Liouville action functional is given by
$$\phi \mapsto \int_{\Sigma} \sqrt{g} \left(\frac{1}{2}g^{ij} \partial_{i} \phi \partial_{j} \phi + R\phi + \exp(\mu \phi) \right )$$
In contrast, the re-normalized Liouville action functional is given by
$$\phi \mapsto \frac{1}{4\pi} \int_{\Sigma}    \sqrt{g} \left  (g^{ij} \partial_{i} \phi \;  \partial_{j} \phi + QR\phi + 4 \pi \mu \exp(2 b \phi) \right )$$
where $b$ is the coupling constant and the background charge $Q$ satisfies $Q=b+b^{-1}$ and is related to the central charge $c_{L}$ of the Liouville theory by
$$c_{L}=1+6Q^{2}$$
These quantities, and in particular the value of $b$, can be related to the central charge $c(p,q)$ of the matter content by an anomaly cancellation calculation and one obtains
$$c_{L}+c(p,q)=26$$
Hence 
$$Q=(4+\frac{(p-q)^{2}}{pq})^{1/2}$$
This means that the coupling constant $b$ satisfies the quadratic equation
$$b^{2}-\frac{p+q}{\sqrt{pq}} \cdot b+1=0$$
and the two roots are
$$b_{1}=\left ( \frac{q}{p} \right )^{1/2}$$
$$b_{2} = \left ( \frac{p}{q} \right )^{1/2}$$
The duality of the $(p,q)$ minimal model coupled to gravity can be interpreted as an invariance of the theory with respect of choosing either one of the two values of the coupling constant. Note however that the resulting theories are not identical but rather dual, meaning that one can be expressed in terms of the other one. There is hence an expected duality 
$$b \longmapsto b^{-1}$$ 
that corresponds to the switch of $p$ and $q$ and hence is sometimes called the p -- q duality. Due to the way the coupling constant enters the Lagrangian this is also called the T -- duality of 2D quantum gravity, see \cite{CIY}. 

In the discretized framework one can rigorously prove this duality, as was done first by Fukuma-Kawai-Nakayam \cite{FKN} and Kharchev-Marshakov \cite{KM} in the early 90's. 

To make the relation with the numerical local Langlands duality, one has to recast the initial formulations, both on the arithmetic and the quantum side: Koch's direct approach to the Langlands duality had to be modified by Laumon by introducing the local $\ell$-adic Fourier transforms. On the physics side, the previous approaches had to be modified in order to phrase the duality in terms of D-modules, as was done by Schwarz and the author in \cite{LUU}, \cite{LUU2}. To put it differently, in order to be able to switch from the arithmetic duality to the quantum duality one should look at the underlying reason that the arithmetic numerical local Langlands correspondence holds: It comes from the symmetries of local Langlands parameters coming from the local Fourier transform. This transform has complex and $\ell$-adic incarnations and this gives the bridge between the dualities. 

For every point $x$ of $\PP^{1}$ there exists an arithmetic local Fourier transform $\mathcal F^{(x,\infty)}_{\textrm{arith}}$ and a geometric version $\mathcal F^{(x,\infty)}_{\textrm{geom}}$.
On the geometric side, these transforms relate connections on various formal punctured discs on the Riemann sphere. For example, the geometric $\mathcal F^{(\infty,\infty)}$ transform is described in the following picture:

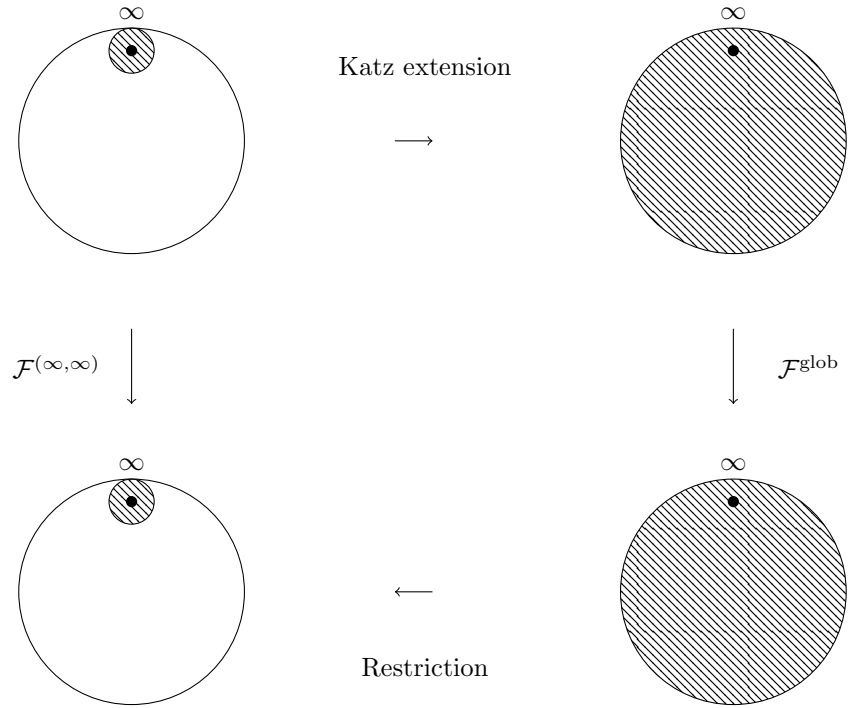
\begin{figure}[H]
\label{Fourier-figure}
\centering
\begin{tikzpicture}
\draw (2,2) circle (1.5cm);
\draw[pattern=north west lines] (2, 3.2) circle (0.3 cm);
\draw (2,3.2) node[circle, inner sep=1.5pt,fill] {} ;
\draw (2,3.7) node {$\infty$};
\draw [->] (5.5,2) -- (6,2);
\draw (5.9,3) node {$\textrm{Katz extension}$};
\draw[pattern=north west lines] (10,2) circle (1.5cm);
\draw (10,3.2) node[circle, inner sep=1.5pt,fill] {} ;
\draw (10,3.7) node {$\infty$};
\draw [->] (10,-0.5) -- (10,-1.5);
\draw [->] (2,-0.5) -- (2,-1.5);
\draw (1,-1) node {$\mathcal F^{(\infty,\infty)}$};
\draw (11,-1) node {$\mathcal F^{\textrm{glob}}$};
\draw[pattern=north west lines] (10,-4) circle (1.5cm);
\draw (10,-2.8) node[circle, inner sep=1.5pt,fill] {} ;
\draw (10,-2.3) node {$\infty$};
\draw [->] (6,-4) -- (5.5,-4);
\draw (5.9,-5) node {$\textrm{Restriction}$};
\draw (2,-4) circle (1.5cm);
\draw[pattern=north west lines] (2, -2.8) circle (0.3 cm);
\draw (2,-2.8) node[circle, inner sep=1.5pt,fill] {} ;
\draw (2,-2.3) node {$\infty$};
\end{tikzpicture}
\caption{The local Fourier transform $\mathcal F^{(\infty,\infty)}$ and its relation to the global  Fourier transform $\mathcal F^{\textrm{glob}}$ of D-modules on the plane $\PP^{1}\backslash \{\infty\}$.}
\end{figure}

In the following statements we suppress the subscripts for local Langlands parameters indicating whether they are arithmetic or geometric. The reason is that one obtains the same statement in both cases: The $\mathcal F^{(0,\infty)}$ local Fourier transform gives rise to a map
$$ \textrm{LL}(n,j) \longrightarrow \textrm{LL}(j+n,j)$$
and for $j>n$ the $\mathcal F^{(\infty,\infty)}$ local Fourier transform gives rise to a map
$$\textrm{LL}(n,j) \longrightarrow \textrm{LL}(j-n,j)$$
See \cite{LAU} (Th\'{e}or\`{e}me 2.4.3) for the arithmetic case and \cite{BE} (Proposition 3.14) for the geometric case. These symmetries are very powerful:

On the arithmetic side, Henniart \cite{HEN} used them to prove the numerical local Langlands correspondence. First he reduces to the case of local fields of positive characteristic $p$ and the case that $n=p^{r}$ and $r$ bigger than the power of $p$ dividing $j$. In this case one immediately obtains the proof: The properties of the transform $\mathcal F^{(0,\infty)}$ imply the symmetry
$$c(n,j)_{\textrm{arith}}-c(n,j-1)_{\textrm{arith}}=c(n+j,j)_{\textrm{arith}}-c(n+j,j-1)_{\textrm{arith}}$$
A double induction on $j$ and the power of $p$ dividing $n$ then shows that
$$c(n,j)_{\textrm{arith}}=(q-1)(q^{j}-q^{j-1}+q^{j-1})=(q-1)q^{j},$$
as desired.

This amazingly concise argument should be contrasted with Koch's \cite{KOC} laborious explicit computations that could only treat the case where the residue characteristic $p$ divides $n$ at most linearly. Laumon's approach via the Fourier transform is completely different: 

The transform gives rise to a hidden symmetry of local Langlands parameters that allows to count them without having an explicit handle on them. One can sum up Henniart's result as follows: For $p$ a positive integer and $q \ge -p$, the map
$$\mathcal F^{(0,\infty)}_{\textrm{arith}}: \textrm{LL}_{\textrm{arith}}(p, p+q)\longrightarrow \textrm{LL}
_{\textrm{arith}}(2p+q, p+q)$$
of arithmetic Langlands parameters gives rise to the arithmetic numerical local Langlands duality. 

Concerning the T -- duality, it follows from our work \cite{LUU}, \cite{LUU2} that when translated to the geometric setting one obtains the duality of 2D quantum gravity. Namely, for $p$ and $q$ positive co-prime integers, the $(p,q)$ model of 2D quantum gravity can be described via the local geometric Langlands parameter $\mathcal L_{p,q}$ described earlier in such a way that the map
$$\mathcal F^{(\infty,\infty)}_{\textrm{geom}} : 
\textrm{LL}_{\textrm{geom}}(p, p+q) \longrightarrow \textrm{LL}_{\textrm{geom}}(q, p+q)$$
realizes the T -- duality of 2D quantum gravity. We refer to \cite{LUU2} for a detailed statement involving control of the dynamics of the KP flows. In particular, $\mathcal L_{p,q}$ does not get mapped to $\mathcal L_{q,p}$ but to a ``time-reversed'' version $\mathcal L_{q,p}'$. 

A unified treatment of the above arithmetic and quantum dualities emerges. The local Fourier transforms, in their $\ell$-adic and complex version, are a source of symmetries of local Langlands parameters. In the arithmetic setting these symmetries imply the numerical local Langlands duality and in the geometric setting they allow to describe the T -- duality:
$$$$
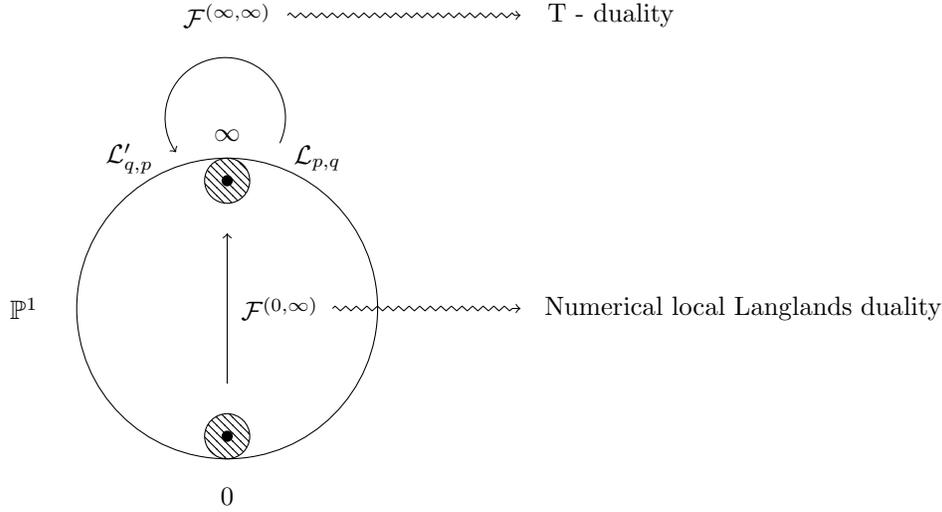
\begin{tikzpicture}
\draw[pattern=north west lines] (2, 0.3) circle (0.3 cm);
\draw  (2,2) circle (2cm);
\draw[pattern=north west lines] (2, 3.7) circle (0.3 cm);
\draw[->] (2,1) --(2, 3) ;
\draw (2.7,2) node {$\mathcal F^{(0,\infty)}$};
\draw (-0.7,2) node {$\PP^{1}$};
\draw (8.8,2) node {$\textrm{ Numerical local Langlands duality}$};
\draw[->] (2.7,4.2) arc (-25 : 215 : 0.8 cm) ;
\draw (2,5.9) node {$\mathcal F^{(\infty,\infty)}$};
\draw (7.1 ,5.9) node {$\textrm{ T - duality }$};
\draw (2,0.3) node[circle, inner sep=1.5pt,fill] {} ;
\draw (2,3.7) node[circle, inner sep=1.5pt,fill] {} ;
\draw (3.2,4) node {$\mathcal L_{p,q}$};
\draw (0.7,4) node {$\mathcal L_{q,p}'$};
\draw (2,-0.5) node {$0$};
\draw (2,4.3) node {$\infty$};
\draw [->,
line join=round,
decorate, decoration={
    zigzag,
    segment length=4,
    amplitude=.9,post=lineto,
    post length=2pt
}]  (3.4,2) -- (5.9,2);
\draw [->,
line join=round,
decorate, decoration={
    zigzag,
    segment length=4,
    amplitude=.9,post=lineto,
    post length=2pt
}]  (2.8,5.9) -- (5.9,5.9);
\end{tikzpicture}

In this sense, the two dualities are simply specific incarnations of the same symmetry principle.

$$$$

\textbf{Acknowledgments}: It is a great pleasure to thank Andrei Jorza, Albert Schwarz, Zhiwei Yun for helpful exchanges.

\Addresses

\end{document}